\documentclass[twocolumn,aps]{revtex4}

\usepackage{amssymb}
\usepackage{graphicx}
\usepackage{dcolumn}
\usepackage{bm}
\usepackage{epsfig}

\begin{document}
\title{Isotropic-nematic phase behavior of length polydisperse hard rods}
\author{H.H. Wensink, and G.J. Vroege}
\email{G.J.Vroege@chem.uu.nl}
\affiliation{Van 't Hoff Laboratory for Physical and Colloid Chemistry,
Debye Institute, Utrecht University,
Padualaan 8, 3584 CH Utrecht, The Netherlands}
\date{\today}

\begin{abstract}
The isotropic-nematic phase behavior of length polydisperse hard rods with
arbitrary length distributions is calculated. Within a numerical treatment
of the polydisperse Onsager model using the Gaussian trial function Ansatz
we determine the onset of isotropic-nematic phase separation, coming from a
dilute isotropic phase and a dense nematic phase. We focus on parent systems
whose lengths can be described by either a Schulz or a `fat-tailed'
log-normal distribution with appropriate lower and upper cutoff lengths. In
both cases, very strong fractionation effects are observed for parent
polydispersities larger than roughly 50 \%. In these regimes, the isotropic
and nematic phases are completely dominated by respectively the shortest and
the longest rods in the system. Moreover, for the log-normal case, we
predict triphasic isotropic-nematic-nematic equilibria to occur above a
certain threshold polydispersity. By investigating the properties of the
coexisting phases across the coexistence region for a particular set of
cutoff lengths we explicitly show that the region of stable triphasic
equilibria does not extend up to very large parent polydispersities but
closes off at a consolute point located not far above the threshold
polydispersity. The experimental relevance of the phenomenon is discussed.
\end{abstract}

\maketitle

\section{Introduction}

Dispersions of highly anisometric rod- or platelike colloidal particles
exceeding a certain concentration are known to undergo a
transition from an isotropic state (I), in which the particles are oriented
in random directions to an aligned nematic state (N) \cite
{Zocher,Langmuir,Bernalnature}. The theoretical basis for this phenomenon
has been established by Onsager in the 1940s. In a seminal paper \cite
{Onsager}, he showed that the phase transition can be explained on the basis
of purely repulsive interactions between the particles. The basic mechanism
behind the transition is a competition between orientional entropy which
favors the isotropic state and the entropy effect associated with the
orientation-dependent excluded volume of the anisometrical particles which
favors the ordered nematic.

A main characteristic of systems of (anisometric) colloidal
particles is their inherent
polydispersity, i.e. the particles may differ in size and shape \cite
{Wierenga,Buiningwater}. The issue of polydispersity and its effect on the
interpretation of experimental results has already been addressed by Onsager
in his original paper. Later on, extensions of the Onsager theory allowing
for phase diagram calculations for bidisperse \cite
{Lekkerkerker84,OdijkLekkerkerker,birshtein,LekkerVroeg,vanroij96/2,vanroij98}
and tridisperse systems \cite{LekkerVroegtridisperse} of hard rods as well as
binary mixtures of hard platelets \cite{wensinkbidikte} revealed a rich
variety in behavior, most notably a widening of the coexistence region, a
fractionation effect (i.e. segregation of the species among the coexisting
phases), a reentrant phenomenon and, most interestingly, the possibility of
a demixing of the nematic phase which may give rise to
isotropic-nematic-nematic triphasic equilibria.

So far, very few theoretical attempts have been made to study the
isotropic-nematic phase behavior of truly polydisperse systems, i.e. systems
characterized by a continuous distribution in particle size, within the
Onsager treatment. As to rodlike systems, there are some studies in which
the effect of polydispersity on the isotropic-nematic transition is
accounted for using perturbation theories \cite{sluckin,chen}. The major
drawback of this approach, in which the effect of polydispersity is
considered as a small perturbation to a monodisperse reference system, is
its limited applicability; it is only justified for slightly polydisperse
systems with very narrow length distributions. Although the perturbation
approach qualitatively predicts some generic features such as a broadening
of the coexistence area and a fractionation effect, with the longest rod
going preferentially into the nematic phase, other interesting phenomena
which are expected to occur at much higher polydispersities (in particular
polydispersity-induced demixing transitions) are not treated and therefore
remain elusive.

Solving the phase equilibrium conditions for systems with
arbitrary size distributions is by no means trivial and requires
considerable numerical effort. In particular, the presence of almost
infinitely many components in a polydisperse system requires an equally
large number of coexistence conditions to be solved simultaneously which
obviously is a formidable task. Recently, a number of studies have appeared in which
the fully polydisperse Onsager model, albeit in simplified form, was
subjected to a numerical treatment. Speranza and Sollich \cite
{sollichonsagerP2,sollichonsagerP2fattail} investigated the model using the
so-called ${\cal P}_{2}$-approximation which consists of truncating the
angular dependence of the orientation-dependent excluded-volume term after
the first nontrivial term in an expansion in terms of even Legendre polynomials ${\cal P}_{2n}$.
A remarkable outcome of these calculations is
that for rod length distributions with sufficiently fat tails (e.g.
log--normal distributions) triphasic isotropic-nematic-nematic equilibria
are predicted to occur in a small interval of polydispersities of the parent
system. However, the simplified orientation distribution function (ODF) for
the nematic phase pertaining to the `${\cal P}_{2}$-model' is only valid for
the description of very weakly aligned nematic phases. The behavior
predicted from this model should therefore be considered with some care,
particularly in those regions where the fractionation effect is strong and
the phase behavior is dominated by the effect of the longest rods in the system.
The presence of very long rods in a nematic state may force the entire
system into a strongly aligned nematic configuration, so that a more
appropriate form for the nematic ODF is required in these cases.

In this paper we use the Gaussian trial ODF approach to calculate the
isotropic-nematic phase behavior of hard rod systems which can be described
by either a Schulz or a log-normal length distributions with arbitrary
polydispersities. The benefit of using the Gaussian Ansatz is twofold.
First, all necessary integrals for the monodisperse Onsager model are
analytically tractable so that only numerical integrations over the length
distributions need to be considered for the polydisperse case. Second, the
Gaussian ODF allows for a qualitatively better description of highly ordered
nematic states compared to the ${\cal P}_{2}$-approximation which makes it a
suitable tool for describing polydisperse systems, particularly the ones
with a `fat-tailed' length distribution.  While work on this subject was
still in progress, Speranza and Sollich reported a numerical analysis of the
exact Onsager model \cite{sollichonsagerexact}, i.e. using the numerically
{\em exact} ODF. Also there, triphasic equilibria were predicted for both
Schulz-type parent distributions and `fat-tailed' log-normal forms. However,
due to the numerical complexity of the problem only the {\em onset} of
nematic ordering from  an isotropic reference phase was considered there so
that no information could be obtained about the properties of the isotropic
and nematic phases across the coexistence region. Consequently, no
conclusive insight could be gained as to whether the triphasic equilibria
constitute a significant part of the phase diagram. Within the Gaussian
approach it is possible to access the coexistence region with only limited
additional numerical effort.  An important consequence is that it also
enables us to study the triphasic demixing in more detail and to gain
insight in the extent of its stability region. Although we cannot calculate
the binodal curves for these equilibria (which locate the precise onset), we
are able to localize the spinodal points for the coexisting nematic phase
across the two-phase region. The presence of these spinodal points indicates
a local instability of the nematic phase which implies that a triphasic
demixing transition must take place.

This paper is structured as follows. In Sec. II we describe the Onsager
theory generalized to polydisperse systems. The conditions for phase
equilibria are outlined in general terms in Sec. III-A, and specified to the
onset of isotropic-nematic phase separation and the full phase split
situation in Secs. III-B and III-C respectively. In III-D some details of
the truncated parent distributions we used in our calculations are given.
The numerical results for the different parent distributions will be
presented and discussed in detail in Secs. IV and V. Finally, we summarize
and conclude in Sec. VI. Technical aspects of the subject are treated in
several appendices. In App. A we provide details about the numerical
procedure we adopted to solve the coexistence conditions. In App. B we
explicitly show that the Gaussian Ansatz yields the exact scaling results
for parent distributions with infinite cutoff lengths and in Sec. C we
establish a spinodal instability criterion for the nematic phase inside the
coexistence region which is used to detect possible
isotropic-nematic-nematic triphasic equilibria along the coexistence
trajectory.

\section{Polydisperse Onsager theory; starting equations}

Let us consider a system of hard rodlike cylinders with equal diameters $D$
but {\em different} lengths $L$ , in a macroscopic volume $V$. \ To
characterize the rod lengths in our polydisperse system we introduce the
relative rod length $l=L/L_{0}$ (with $L_{0}$ some reference rod length $%
L_{0}$) which is assumed to be a {\em continuous} variable. We may then take
the limit $L_{0}/D\rightarrow \infty $ (infinitely thin rods) at constant
values for the relative lengths $l$. In the Onsager approach, the excess
free energy describing the excluded volume interactions between the
particles is truncated after the second virial term. This approach can be
shown to yield the exact free energy in the limit of infinitely thin rods. A
generalization of the original Onsager model to include polydispersity leads
to the following expression for the total Helmholtz free energy density $f$
(in units $k_{B}T\equiv \beta ^{-1}$)

\begin{eqnarray}
f &\equiv &\frac{b\beta F}{V}\sim \int c(l)[\ln c(l)-1]dl+\int c(l)\omega
(l)dl  \nonumber \\
&&+\int \int c(l)c(l^{\prime })ll^{\prime }\rho (l,l^{\prime })dldl^{\prime
}.  \label{freeenergy}
\end{eqnarray}
All irrelevant contributions linear in $c$ arising from the standard
chemical potentials of the particles are omitted since they only depend on
the solvent chemical potential and the temperature. The concentrations $c$
are rendered dimensionless by relating them to the orientationally averaged
excluded volume per particle between two reference rods, $b=\pi DL_{0}^{2}/4$%
, via $c(l)=bN(l)/V$ where $N(l)dl$ is the number of particles with relative
length between $l$ and $l+dl.$ The density distribution over lengths $c(l)$
can be decomposed according to $c(l)=c_{0}p(l)$, with $p(l)$ the normalized
length distribution ($\int dlp(l)\equiv 1$) and $c_{0}$ the total
dimensionless rod concentration.

The free energy Eq. (\ref{freeenergy}) consists of several entropic
contributions. The first term represents the exact ideal free energy of the
polydisperse system. The second term contains parameter $\omega $ as a
measure for (the negative) of the orientational entropy \cite{Onsager}
\begin{equation}
\omega (l)\equiv \int \psi (l,\Omega )\ln [4\pi \psi (l,\Omega )]d\Omega ,
\label{sor}
\end{equation}
where $\psi (\theta ,l)$ is the normalized ODF for species $l$ describing
the distribution of the particles' solid angle $\Omega $. In the isotropic
state, all orientations are equally probable so that $\psi _{\text{iso}}$ is
simply a constant ($1/4\pi $) independent of $l$. In the nematic state,
however, the ODFs\ are peaked functions (generally different for each
species $l$), due to the fact that the rods are aligned along a nematic
director. Note that $\omega $ (and hence the orientation free energy)
attains its minimum ($\omega =0$) in the isotropic state, whereas $\omega >0$
in the nematic state.

The last term in Eq. (\ref{freeenergy}) describes the excess free energy
which accounts for the particle interactions. In a second virial
approximation, the interactions between hard particles may be expressed as
an excluded volume entropy depending on the excluded volume between two
particles. A measure for the average excluded-volume interaction between
rods of relative length $l$ and $l^{\prime }$ is given by the following
angular average
\begin{equation}
\rho (l,l^{\prime })\equiv \frac{4}{\pi }\int \int \left| \sin \gamma
(\Omega ,\Omega ^{\prime })\right| \psi (l,\Omega )\psi (l^{\prime },\Omega
^{\prime })d\Omega d\Omega ^{\prime }.  \label{rhodef}
\end{equation}
Using the isotropic average $\left\langle \left\langle \left| \sin \gamma
(\Omega ,\Omega ^{\prime })\right| \right\rangle \right\rangle =\pi /4$ we
obtain $\rho (l,l^{\prime })\equiv 1$ for the isotropic state. This
indicates that the excluded volume (or packing) free energy is indeed
maximal in the isotropic phase but decreases as soon as the rods align to
form a nematic phase.

In this study we use Gaussian trial ODFs with variational parameter $\alpha
(l)$ to describe the angular distribution of rods with relative length $l$
in the nematic state \cite{OdijkLekkerkerker}. The Gaussian Ansatz consists
of supposing
\begin{equation}
\psi (l,\theta )\equiv \left\{
\begin{tabular}{ll}
$\frac{\alpha (l)}{4\pi }\exp [-\frac{1}{2}\alpha (l)\theta ^{2}]$ & $\qquad
0\leq \theta \leq \frac{\pi }{2}$ \\
&  \\
$\frac{\alpha (l)}{4\pi }\exp [-\frac{1}{2}\alpha (l)(\pi -\theta )^{2}]$ & $%
\qquad \frac{\pi }{2}\leq \theta \leq \pi $%
\end{tabular}
\right. ,  \label{ODF}
\end{equation}
where $\alpha $ is now a function of $\ l$. Note that, due to the uniaxial
symmetry of the nematic phase, the ODFs only depends upon the polar angle $%
\theta $ between the particle orientation vector and the nematic director.
Inserting Eq. (\ref{ODF}) in Eq. (\ref{sor}) and straightforward integration
yields for the orientational entropy
\begin{equation}
\omega (l)\sim \ln \alpha (l)-1.\qquad \qquad  \label{sorex}
\end{equation}
For the excluded volume entropy in the nematic phase $\rho _{\text{nem}%
}(l,l^{\prime })$ only the leading order term of its asymptotic expansion
for large $\alpha $ will be retained
\begin{equation}
\rho _{\text{nem}}(l,l^{\prime })\sim \sqrt{\frac{8}{\pi }\left( \frac{1}{%
\alpha (l)}+\frac{1}{\alpha (l^{\prime })}\right) }+{\cal O}\left[ \alpha
^{-3/2}(l),\alpha ^{-3/2}(l^{\prime })\right].  \label{rhonemalfa12}
\end{equation}
Substituting Eqs. (\ref{sorex}) and (\ref{rhonemalfa12}) into Eq. (\ref
{freeenergy}) and minimizing the free energy density with respect to the
non-conserved orientational degrees of freedom by means of a functional
differentiation with respect to $\alpha (l)$ gives
\begin{eqnarray}
\frac{\delta f}{\delta \alpha (l)}\sim &&\frac{c(l)}{\alpha (l)}-\left( \frac{8%
}{\pi }\right) ^{1/2}\frac{lc(l)}{2\alpha ^{2}(l)}   \nonumber \\
&&\times\int l^{\prime}c(l^{\prime })\left( \frac{1}{\alpha (l)}+\frac{1}{\alpha (l^{\prime })}%
\right) ^{-1/2}dl^{\prime }.
\end{eqnarray}
Applying the stationarity condition $\delta f/\delta \alpha (l)\equiv 0$ and
some rearranging leads to the following self-consistency equation
\begin{equation}
\tilde{\alpha}(l)=2l^{2}\left\{ \int l^{\prime }p^{(N)}(l^{\prime })\left[ 1+%
\frac{\tilde{\alpha}(l)}{\tilde{\alpha}(l^{\prime })}\right]
^{-1/2}dl^{\prime }\right\} ^{2},  \label{selfconsistency}
\end{equation}
Here, we have factorized the Gaussian variational parameter function $\alpha
(l)$ into a concentration-dependent part and a contribution $\tilde{\alpha}%
(l)$ only related, via Eq.(\ref{selfconsistency}), to the normalized length
distribution in the nematic phase $p^{(N)}(l)$. Hence we write
\begin{equation}
\alpha (l)=\tilde{\alpha}(l)\frac{4c_{0}^{2}}{\pi }.
\end{equation}
showing that for all $l$ the variational parameter $\alpha $ depends
quadratically on $c_{0}$ just as in the monodisperse case \cite{Vroege92}. An
approximate analytical solution to Eq. (\ref{selfconsistency}) valid for
infinitely narrow distributions \cite{sluckin} (denoted by subscript $\delta $%
) can be obtained by substituting a delta function $p^{(N)}(l)=\delta (l-1)$
which gives
\begin{equation}
\tilde{\alpha}_{\delta }(l)=\frac{1}{2}\left( \sqrt{8l^{2}+1}-1\right) .
\label{alphanarrow}
\end{equation}
This result may be interpreted as a measure for the nematic alignment of a
{\em single} rod with relative length $l$ added to a nematic bulk system of
monodisperse rods with reference length $L_{0}$. Eq. (\ref{alphanarrow})
shows that $\tilde{\alpha}_{\delta }(l)$ and hence the order parameter \cite
{Vroege92} $P(l)\sim 1-3/\alpha (l)$ are in general, as we might have
anticipated, increasing functions of the relative rod length, i.e. $\tilde{%
\alpha}_{\delta }(l)\propto l$ for large $l$. Moreover, $\tilde{\alpha}%
_{\delta }(0)=0$ which means that there is no ordering for rods of zero
length, as formally must be the case. However, it should be pointed out
that rods with lengths close to zero must be excluded from our model because
the normalization factors for the Gaussian ODFs in Eq. (\ref{ODF}) do not
allow for a correct description of isotropically distributed or weakly
aligned species in the nematic state. For consistency reasons we must
therefore introduce a lower limit ($l_{\min }>0$) in all length
distributions.

\section{I-N phase coexistence}

\subsection{Equilibrium conditions for polydisperse systems}

The conditions for phase equilibrium are that the coexisting isotropic and
nematic phases must have equal chemical potential $\mu (l)$ for all relative
rod lengths $l$, as well as equal osmotic pressure $\Pi $. The chemical
potential can be derived by functional differentiation of the free energy
with respect to the length distribution $c(l)$%
\begin{equation}
\beta \mu (l)=\frac{\delta f}{\delta c(l)}.  \label{defmu}
\end{equation}
Using Eqs. (\ref{sorex}) and (\ref{rhonemalfa12}) together with the
isotropic values $\omega \equiv 0$ and $\rho \equiv 1$ we obtain
\begin{eqnarray}
\beta \mu _{\text{iso}}(l) &=&\ln c^{(I)}(l)+2lc_{1}^{(I)}  \nonumber \\
\beta \mu _{\text{nem}}(l) &=&\ln c^{(N)}(l)+\ln \left[ \frac{4}{\pi }%
(c_{0}^{(N)})^{2}\tilde{\alpha}(l)\right] -1 \nonumber \\
&& +\tilde{\mu}_{\text{ex}}^{(N)}(l)  \label{chempot},
\end{eqnarray}
where $c_{1}$ denotes the first {\em moment density} following from the
definition
\begin{equation}
c_{k}=c_{0}m_{k}=\int dll^{k}c(l),\qquad k=0,1,2,\ldots
\end{equation}
Here, $m_{k}$ denotes the $k$-th moment of the (normalized) distribution.
The excess chemical potential for the nematic phase $\tilde{\mu}_{\text{ex}%
}^{(N)}(l)$ is given by
\begin{equation}
\tilde{\mu}_{\text{ex}}^{(N)}(l)=\bigskip 2^{3/2}l\int dl^{\prime
}p^{(N)}(l^{\prime })l^{\prime }\left( \frac{1}{\tilde{\alpha}(l)}+\frac{1}{%
\tilde{\alpha}(l^{\prime })}\right) ^{1/2}.  \label{haa}
\end{equation}
and is independent of the concentration of the nematic phase. Similarly to
Eq. (\ref{selfconsistency}) we can straightforwardly obtain an analytical
solution for $\tilde{\mu}_{\text{ex}}^{(N)}(l)$ valid for near monodisperse
distributions by substituting $p^{(N)}(l)=\delta (l-1)$, which yields the
following scaling result
\begin{equation}
\ \tilde{\mu}_{\text{ex,}\delta }^{(N)}(l)\propto l\sqrt{\frac{1}{\tilde{%
\alpha}_{\delta }(l)}+1} .  \label{muexnarrow}
\end{equation}
Using Eq. (\ref{alphanarrow}) it follows that $\ \tilde{\mu}_{\text{ex}%
}^{(N)}(l)\propto l,$ for very large $l.$ This means that the excess
chemical potential (i.e. the reversibel work required to insert a single rod
of length $l$ in a nematic system of reference rods ) increases with length,
which is consistent with intuition. The osmotic pressure can be written in
terms of the chemical potential and the free energy via
\begin{equation}
b\beta \Pi \equiv -f+\beta \int dlc(l)\mu (l),  \label{pressure}
\end{equation}
which immediately yields for the isotropic phase
\begin{equation}
b\beta \Pi _{\text{iso}}\sim c_{0}^{(I)}+(c_{1}^{(I)})^{2}.
\label{osmpressures}
\end{equation}
For the nematic phase Eq. (\ref{pressure}) this formally gives
\begin{equation}
b\beta \Pi _{\text{nem}}\sim c_{0}^{(N)}+f_{\text{ex}}^{(N)}.
\end{equation}
However, this result can be simplified considerably by noting that the
volume fraction of the average excluded volume (per particle) in the nematic
phase is a constant, namely
\begin{equation}
N\frac{\left\langle \left\langle V_{\text{excl}}\right\rangle \right\rangle
_{l,l^{\prime }}}{V}\sim c_{0}^{(N)}\left\langle \left\langle ll^{\prime
}\rho (l,l^{\prime })\right\rangle \right\rangle _{l,l^{\prime }}=2.
\end{equation}
The brackets denote averages over the normalized length distribution. This
result, which is due to Odijk \cite{odijkLC}, generally holds for both
monodisperse and polydisperse systems, independent of their composition.
From the free energy Eq. (\ref{freeenergy}) it then follows that $f_{\text{ex%
}}^{(N)}=2c_{0}^{(N)}$ so that the osmotic pressure of the nematic phase
reduces to
\begin{equation}
b\beta \Pi _{\text{nem}}\sim 3c_{0}^{(N)},  \label{driecee}
\end{equation}
like for a monodisperse system \cite{Vroege92}. We can now state the
conditions for the coexistence between the isotropic and nematic daughter
phases into which a parent phase (henceforth denoted with superscript 0)
with length distribution $c^{(0)}(l)$ is assumed to have split. From Eq. (%
\ref{chempot}), equality of chemical potentials of both phases is obeyed
exactly if the distributions in the phases have the following form
\begin{equation}
c^{(a)}(l)=W(l)\exp [\xi ^{(a)}(l)],\qquad \qquad a=I,N
\end{equation}
where $W(l)\equiv \exp [\beta \mu (l)]$ must be a function common to both
phases, since $\mu ^{(I)}(l)=\mu ^{(N)}(l)=\mu (l)$. The functions $\xi
^{(a)}(l)$ are given by
\begin{eqnarray}
\xi ^{(I)}(l) &=&-2lc_{1}^{(I)}  \nonumber \\
\xi ^{(N)}(l) &=&\left( 1-\ln \frac{4}{\pi }\right) -2\ln c_{0}^{(N)}-\ln
\tilde{\alpha}(l)-\tilde{\mu}_{\text{ex}}^{(N)}(l). \nonumber \\
\label{xiparameters}
\end{eqnarray}
Furthermore, conservation of matter requires
\begin{equation}
c^{(0)}(l)=(1-\gamma )c^{(I)}(l)+\gamma c^{(N)}(l),  \label{consmatter}
\end{equation}
where $\gamma $ denotes the fraction of the system volume occupied by the
nematic phase. Using Eq. (\ref{consmatter}), we can express $W(l)$  in
terms of the parent distribution $c^{(0)}(l)$ which gives
\begin{eqnarray}
c^{(a)}(l) &=&c^{(0)}(l)\frac{\exp [\xi ^{(a)}(l)]}{(1-\gamma )\exp [\xi
^{(I)}(l)]+\gamma \exp [\xi ^{(N)}(l)]}. \nonumber \\ \label{eqdistr}
\end{eqnarray}
These functions represent the equilibrium rod length distributions for the
coexisting phases. The phase equilibria can now, in principle, be obtained
by solving a set of self-consistency equations for the moment densities of
both phases and for the functions $\tilde{\alpha}(l)$ and $\tilde{\mu}_{%
\text{ex}}^{(N)}(l)$ pertaining to\ the nematic phase. These equations will
be worked out below for a specific situation, namely at the onset of
isotropic-nematic phase separation.

\subsection{The onset of I-N phase separation; cloud and shadow curves}

In this section we aim at locating the onset of isotropic-nematic phase
separation indicated by so-called cloud and shadow points. A cloud point
marks the density where a parent phase starts to split off an infinitesimal
amount of a new coexisting phase, called the shadow phase. Accordingly, at
the isotropic cloud point only an infinitesimal amount of nematic phase
(shadow phase) has emerged and so the distribution of the isotropic phase is
only negligibly perturbed away from the parent. Hence, for the isotropic
cloud point we may set $\gamma =0$ in Eq. (\ref{eqdistr}) so that,
\begin{equation}
c^{(I)}(l)=c^{(0)}(l),
\end{equation}
implying that the distribution in the isotropic phase at the cloud point is
equal to the parent distribution and hence $c_{0}^{(I)}=c_{0}^{(0)}$. The
(normalized) rod distribution in the nematic shadow phase (with density $%
c_{0}^{(N)}$) is now given by
\begin{eqnarray}
p^{(N)}(l) &=&\frac{c^{(0)}(l)}{c_{0}^{(N)}}\exp \left[ \xi ^{(N)}(l)-\xi
^{(I)}(l)\right]   \nonumber \\
&=&{\cal K}_{N}\frac{p^{(0)}(l)}{\tilde{\alpha}(l)}\exp \left[ 2c_{0}^{(0)}l-%
\tilde{\mu}_{\text{ex}}^{(N)}(l)\right] ,  \label{exprnemshadow}
\end{eqnarray}
where ${\cal K}_{N}=$ $\pi ec_{0}^{(0)}/4(c_{0}^{(N)})^{3}$ and $p^{(0)}(l)$
the normalized parent distribution. Note that Eq. (\ref{exprnemshadow}) is
an implicit expression for $p^{(N)}(l)$ because it still depends on the
unknown functions for the variational parameter \ $\tilde{\alpha}(l)$ and
the excess \ chemical potential $\tilde{\mu}_{\text{ex}}^{(N)}(l)$ for each
species in the nematic shadow phase. Explicit solutions for these functions
can be obtained by substituting Eq. (\ref{exprnemshadow}) into Eqs. (\ref
{selfconsistency}) and (\ref{haa}) and numerically solving the resulting
self-consistency equations.

The concentrations of the isotropic cloud phase and the coexisting nematic
shadow are found by imposing the normalization condition for the
distribution in the nematic shadow phase,
\begin{equation}
\int p^{(N)}(l)\equiv 1,  \label{normcond}
\end{equation}
and the condition of equal osmotic pressure
\begin{equation}
3c_{0}^{(N)}=c_{0}^{(0)}+(c_{1}^{(0)})^{2}.  \label{condeqpressure}
\end{equation}
Using this simple equation to eliminate e.g. $c_{0}^{(N)},$ we may
conveniently combine Eqs. (\ref{normcond}) and (\ref{condeqpressure}) into
one self-consistency equation for the concentration of the \ isotropic cloud
point, which we can solve in an iterative fashion. However, since $\tilde{%
\alpha}(l)$ and $\tilde{\mu}_{\text{ex}}^{(N)}(l)$ also depend on $%
c_{0}^{(0)}$ [via $p^{(N)}(l),$ Eq. (\ref{exprnemshadow})] this equation has
to be solved along with the coupled set of self-consistency equations, Eqs. (%
\ref{selfconsistency}) and (\ref{haa}), so that we end up with a set of
three coupled nonlinear equations. Obviously, solving this set is not a
trivial task but requires some numerical effort. For this reason we have
devoted an Appendix A to this issue in which we describe some details of the
numerical procedures adopted in this study.

We can now perform a similar analysis to obtain expressions for the {\em %
nematic} cloud point and the associated {\em isotropic} shadow point, which
locate the onset of I-N equilibrium coming from a dense nematic parent
phase. Since the latter now coexists with an infinitesimal amount of an
isotropic shadow phase we may set $\gamma =1$ in (\ref{eqdistr}) so that $%
c^{(N)}(l)=c^{(0)}(l)$ and $c_{0}^{(N)}=c_{0}^{(0)}$. The (normalized) rod
distribution in the isotropic shadow phase (with density $c_{0}^{(I)}$) is
then given by
\begin{eqnarray}
p^{(I)}(l) &=&\frac{c^{(0)}(l)}{c_{0}^{(I)}}\exp \left[ \xi ^{(I)}(l)-\xi
^{(N)}(l)\right]   \nonumber \\
&=&{\cal K}_{I}p^{(0)}(l)\tilde{\alpha}(l)\exp \left[ -2c_{1}^{(I)}l+\tilde{%
\mu}_{\text{ex}}^{(N)}(l)\right] ,  \label{citeratie}
\end{eqnarray}
where ${\cal K}_{I}=4(c_{0}^{(0)})^{3}/\pi ec_{0}^{(I)}.$ Since the
functions $\tilde{\alpha}(l)$ and $\tilde{\mu}_{\text{ex}}^{(N)}(l)$
correspond to the nematic parent phase we may substitute $
p^{(N)}(l)=p^{(0)}(l)$ \ in Eqs. (\ref{selfconsistency}) and (\ref{haa}).
Consequently, as the normalized parent distributions $\ p^{(0)}(l)$ \ have a
predescribed form and do {\em not} depend on any concentration, the coupled
self-consistency equations need to be solved only {\em once }for a given $
p^{(0)}(l)$ . The forms of the parent distributions will be specified in
Sec. III-D.

Once the solutions for $\tilde{\alpha}(l)$ and $\tilde{\mu}_{\text{ex}
}^{(N)}(l)$ have been obtained, the concentrations of the nematic cloud and
shadow point can be calculated by requiring self-consistency for the zeroth
moment (normalization condition) and the first moment of the isotropic
shadow distribution, i.e.
\begin{eqnarray}
\int p^{(I)}(l) &=&1,  \nonumber \\
\int lp^{(I)}(l) &=&m_{1}^{(I)}=\frac{c_{1}^{(I)}}{c_{0}^{(I)}}.
\end{eqnarray}
These conditions can, similarly to the previous case, easily be rewritten
into a set of consistency relations for the densities of the nematic cloud
and shadow points, which we can solve iteratively (see also Appendix A).

\subsection{Inside the coexistence region}

We will now focus on the coexistence region between the isotropic and
nematic cloud points, where both phases coexist in noninfinitesimal amounts,
i.e. $0<\gamma <1.$ According to Eq. (\ref{eqdistr}), the equilibrium length
distributions in the coexisting phases are then given by
\begin{eqnarray}
c^{(I)}(l) &=&\frac{c^{(0)}(l)}{\gamma \exp \left\{ \Delta \xi \lbrack
\tilde{\alpha}(l),\tilde{\mu}_{\text{ex}}^{(N)}(l)]\right\} +(1-\gamma )},
\nonumber \\
c^{(N)}(l) &=&\frac{c^{(0)}(l)}{(1-\gamma )\exp \left\{ -\Delta \xi \lbrack
\tilde{\alpha}(l),\tilde{\mu}_{\text{ex}}^{(N)}(l)]\right\} +\gamma}.
\end{eqnarray}
with\ $\Delta \xi (l)=\xi ^{(N)}(l)-\xi ^{(I)}(l)$, given by Eq. (\ref
{xiparameters}). Note that both distributions are now different from the
parental one. Solving the coexistence problem is done in a similar way to
the one described in Sec. III-B for the isotropic cloud and shadow points.
From an experimental point of view, we are only interested in results
located on so-called physical dilution lines along which the overall system
number density ($c_{0}^{(0)}$) is changed (by e.g. adding or evaporating
solvent) while the overall composition of the species ($p^{(0)}(l)$) remains
fixed. The parent distributions will be specified below.

\subsection{Parent distributions}

The numerical method described in the previous sections allows us to
calculate the isotropic-nematic phase diagram for in principle arbitrary
parent distributions. In our study we specify two types of distributions.
The first one is the Schulz distribution which has the form
\begin{equation}
p^{(0)}(l)=Nl^{z}\exp [-(z+1)l],  \label{schulz}
\end{equation}
with normalization factor $N$. In order to exclude rods with potentially
zero length we truncate the distribution at some lower cutoff length $l_{%
\text{min}}$. Henceforth we fix $l_{\text{min}}=0.01$. For calculational
purposes (see Appendix A) we must also have some finite cutoff length $%
l_{\max }$ at large $l$. Of course, introducing finite cutoff lengths is
also reasonable \ from a physical standpoint. The first and second moment
(defined as $m_{k}=\left\langle l^{k}\right\rangle $, $k=1,2)$ of the Schulz
distribution are $m_{1}=1$ and $m_{2}=(z+2)/(z+1)$ only for the {\em %
unbounded} case. However for finite cutoff lengths the moments will deviate
from these values. Although the corrections are generally small, in
particular for large $l_{\text{max}}$, they cannot be neglected. Therefore,
we choose to calculate all relevant moments of the parent distribution
numerically via $m_{k}=\int_{l_{\text{min}}}^{l_{\text{max}}}l^{k}p^{(0)}dl$%
. The polydispersity $\sigma $ is defined as
\begin{equation}
\sigma ^{2}=\frac{\left\langle l^{2}\right\rangle }{\left\langle
l\right\rangle ^{2}}-1,
\end{equation}
and would yield $\sigma ^{2}=(1+z)^{-1}$ for the unbounded Schulz
distribution.

The second distribution we consider is the log-normal one. The `fat-tailed'
log-normal distribution decays much slower at large $l$ than the Schulz one
and therefore possesses a significantly larger contribution of long rods.
The log-normal distribution reads
\begin{equation}
p^{(0)}(l)=Nl^{-1}\exp \left[ -\left( \frac{\ln l-\mu }{2w^{2}}\right) ^{2}%
\right] .  \label{lognormal}
\end{equation}
For the unbounded log-normal distribution, $w$ is directly related to the
polydispersity via $w^{2}=\ln (1+\sigma ^{2})$ and the parameter $\mu $ is
chosen such that $m_{1}=1$, giving $\mu =-w/2.$ The second moment is then
given by $m_{2}=1+\sigma ^{2}$. Also here, truncation of the distribution at
some finite values $l_{\text{min}}$ and $l_{\text{max}}$ leads to deviations
for which we correct numerically.

\section{Results for the onset of I-N phase separation}

\subsection{Schulz distributions}
In Fig. 1 to 5 we have depicted the results for a Schulz parent distribution
with cutoff lengths $l_{\text{min}}=0.01$ and $l_{\text{max}}=100$. The
curves describing the densities of the cloud and shadow phases are shown in
Figs. 1 and 2.
\begin{figure}
\includegraphics{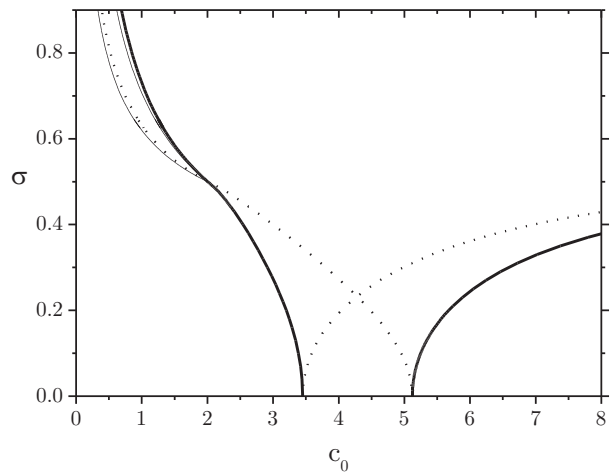}
\caption{Concentrations of the isotropic and nematic cloud
phases (solid lines) and the corresponding shadow phases (dotted lines)
plotted against  (on the vertical axis) the polydispersity $\sigma $ of a
Schulz parent with $l_{\min }=0.01$\ and $l_{\text{max}}=100.$\ The
isotropic cloud curve is the one with the lowest concentration. In the
monodisperse limit ($\sigma =0$) the isotropic cloud point meets the shadow
of the nematic cloud point and vice versa. The thin solid lines are the
limiting curves for $l_{\max }\rightarrow \infty $, given by Eq. (\ref
{limitingcurves}) in Appendix B.}
\end{figure}

\begin{figure}
\includegraphics{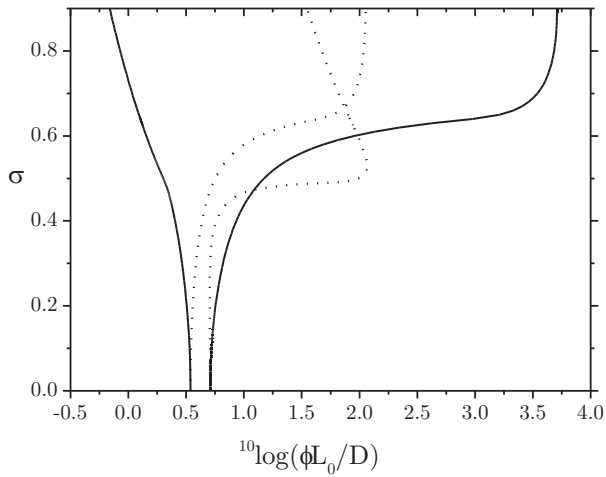}
\caption{Similar to FIG. 1. but with the logarithm of the scaled
volume fraction $\phi L_{0}/D$ plotted versus the parent polydispersity $%
\sigma $ on the vertical axis. Note the dramatic increase of the volume
fraction of the nematic cloud phase.}
\end{figure}

\begin{figure}
\includegraphics{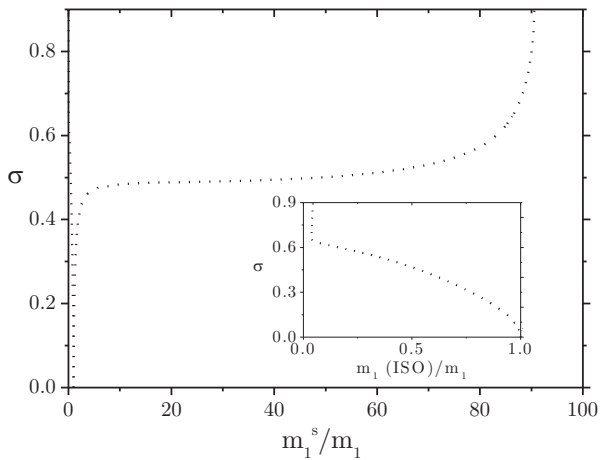}
\caption{Average length $\left\langle l\right\rangle =m_{1}^{s}$
in the isotropic and nematic shadow phases relative to the average length $%
m_{1}$ in the cloud phase plotted against (on the vertical axis) the
polydispersity $\sigma $ of a Schulz parent with $l_{\min }=0.01$ and $l_{%
\text{max}}=100$. The inset shows the relative average length in the
isotropic shadow phase (corresponding to the nematic cloud point).}
\end{figure}
A striking broadening of the coexistence gap can be detected,
mainly due to a dramatic increase of the concentration of the nematic cloud
phase. In Fig. 2 we see that the volume fraction of the nematic cloud phase
increases by several orders of magnitude at $\sigma >0.4$. Although the
nematic shadow curve crosses the corresponding isotropic cloud curve at $%
\sigma \approx 0.5$ in Fig. 1, the volume fraction (and hence the mass
density) of the nematic shadow remains higher than that of the isotropic
cloud phase throughout the phase diagram as we see in Fig. 2. Fig. 3 shows
the extent of fractionation, i.e. the repartitioning of the long and short
rods, among the coexisting phases at the onset of phase separation.
A marked feature is the rapid increase of the average length in the nematic shadow
around some `transitional' polydispersity $\sigma _{t}\simeq 0.5$. This
indicates that the nematic phase becomes preferably populated by the longest
rods in the system. Note that there is a similar effect in the {\em isotropic%
} shadow phase around $\sigma _{t}\simeq 0.7$ where the {\em shortest }rods
completely dominate the isotropic shadow phase at higher polydispersities.
The same effects are reflected somewhat clearer in Fig. 4 showing the
evolution of the polydispersity of the shadow phases. At $\sigma =\sigma
_{t},$ the polydispersity of the shadow phases show a kink. The strong
decrease at higher $\sigma $ is due to the effect that the shadow phases
become more and more enriched in either the longest or the shortest rods in
the distribution. The dramatic change of the composition across the $\sigma $%
-range is shown explicitly in Fig. 5 where we have depicted the normalized
length distributions in the nematic shadow phase for various $\sigma $. A
similar picture is obtained for the distributions in the isotropic shadow
phase (not shown here) but with the peak of the distribution shifting
rapidly towards the lower cutoff length $l_{\text{min}}=0.01$.

\begin{figure}
\includegraphics{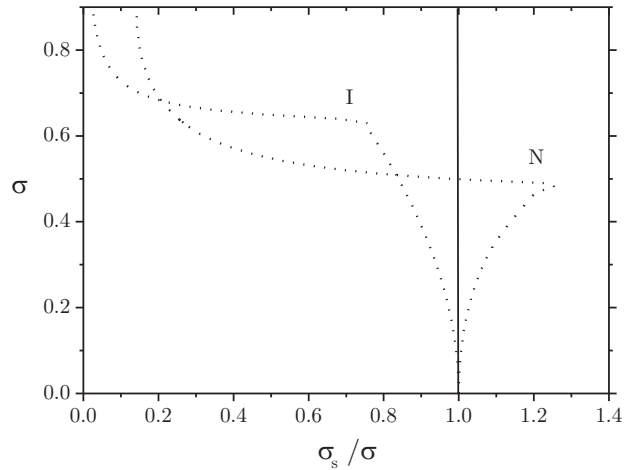}
\caption{Similar to FIG. 3. but with the polydispersity of the
isotropic and nematic shadow phases $\sigma _{s}$ (relative to the parental
one $\sigma )$ plotted against $\sigma $. Note the kinks in the isotropic
and nematic branches around $\sigma =0.65$ and $0.5$, respectively.}
\end{figure}

\begin{figure}
\includegraphics{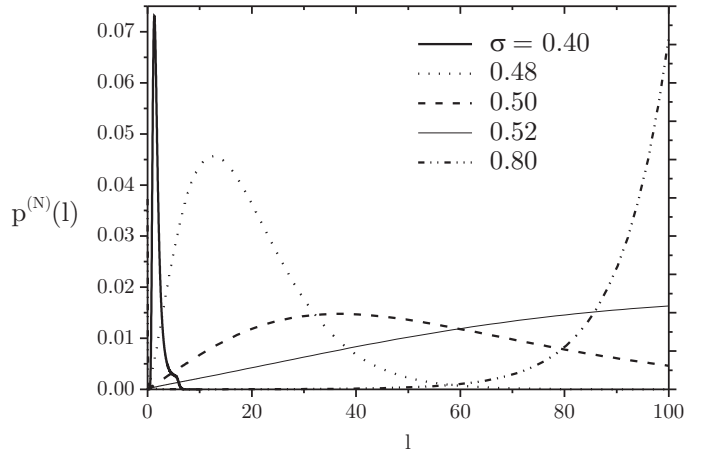}
\caption{Normalized length distributions in the nematic shadow
phases $p^{(N)}(l)$ for various parent polydispersities $\sigma $
of the same Schulz parent as in the previous figures.}
\end{figure}

In summary, we can state that there are two fractionation regimes for the
onset of phase separation. First, at low polydispersities ($\sigma <\sigma
_{t}$)\ moderate fractionation is observed and the shadow phases are mainly
populated by rods with slightly higher (or lower) than average length.
Second, at higher polydispersity ($\sigma >\sigma _{t}$) strong
fractionation occurs such that the shadow phases are completely dominated by
the longest (or shortest) rods in the distribution. In a small interval
around $\sigma =\sigma _{t}$ the location of the peak of the length
distribution shifts rapidly, upon increasing $\sigma ,$\ from a value
slightly different from one (pertaining to the low-$\sigma $ regime) to a
value close to the the cutoff length (corresponding to the high-$\sigma $
regime).

\begin{figure}
\includegraphics{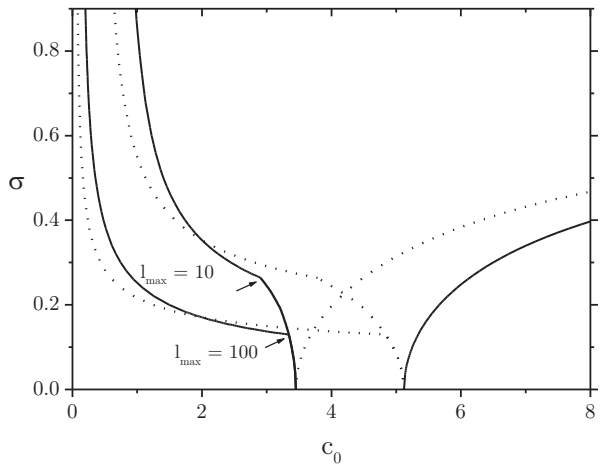}
\caption{Concentrations of the isotropic and nematic cloud
phases (solid lines) and the corresponding shadow phases (dotted lines)
plotted against (on the vertical axis) the polydispersity $\sigma $ of
a log-normal parent distribution with the same lower cutoff length $l_{\min
}=0.01$ but two different higher cutoff lengths, $l_{\text{max}}=10$ and $%
100.$ The isotropic cloud curve is the one with the lowest concentration. The
nematic cloud curve and the associated isotropic shadow curve are
insensitive to the value of $l_{\text{max}}$ on this scale and therefore the results
for $l_{\text{max}}=10$ and $100$ overlap. At the kink in the isotropic cloud curve the isotropic cloud phase
coexists with two nematic phases differing in composition.}
\end{figure}
\begin{figure}
\includegraphics{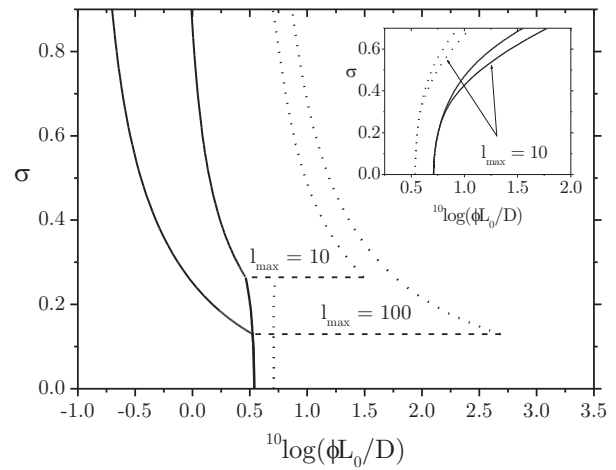}
\caption{Similar to FIG. 6. but with the logarithm of the scaled
volume fraction $\phi L_{0}/D$ plotted versus the parent polydispersity $%
\sigma $ on the vertical axis. The main graph shows the results for the
isotropic cloud and the nematic shadow phase. Note that the jump in the shadow curve
corresponds to a kink in the associated cloud curve.  The inset shows
the results for the nematic cloud and its isotropic shadow phase.}
\end{figure}

So far, we have only discussed the results for a single set of cutoff
lengths. Although the results for different cutoff lengths can be
significantly different, particularly in the `cutoff-dependent' regime \cite
{sollichonsagerP2fattail} $\sigma >\sigma _{t}$, the global phase behavior
remains qualitatively the same. Therefore we conclude that the
aforementioned fractionation scenario generally holds for any Schulz parent
with sufficiently extreme cutoff lengths ( $l_{\min }\ll 1$ and $l_{\text{max%
}}\gg 1$). An interesting limiting case however is the behavior for Schulz
parents with infinitely large cutoff lengths. For this specific case, we
could obtain simple scaling relations which describe the global behavior of
the nematic shadow in the limit of an unbounded Schulz parent, i.e. $l_{%
\text{max}}\rightarrow \infty $. The scaling analysis, worked out briefly in
Appendix B, is closely related to a more elaborate analysis presented in
Ref. \cite{sollichonsagerexact} for the exact Onsager model. In particular,
we show in Appendix B that the Gaussian Ansatz must yield the {\em exact}
high-cutoff scaling relations. The reason for this is that our high-cutoff
scaling form for $p^{(N)}(l)$ (for the nematic shadow phase) is completely
analogous to the exact scaling form.

Finally, we remark that we do not observe a real jump in the shadow curves
(and a kink in the associated cloud curves), as found in Ref. \cite
{sollichonsagerexact}. The presence of a jump in the shadow curve indicates
that, at some point, the isotropic cloud phase coexists with two different
nematic shadow phases and that a region of stable triphasic equilibria
developes. Therefore we conclude that, within the Gaussian Ansatz, the
Schulz form does not give rise to a three-phase separation, at least up to $%
l_{\text{max}}=100$.

\subsection{Log-normal distributions}

The results for the log-normal case are presented in Fig. 6 to 8. The cloud
and shadow curves shown in these figures correspond to log-normal parent
distributions with the same lower cutoff length $l_{\text{min}}=0.01$ but
two different higher cutoffs $l_{\text{max}}=10$ and $100$. We see that the
phase behavior is globally the same as for the Schulz case; there is a
generic broadening of the biphasic region (Fig. 7) and a very pronounced
fractionation effect, particularly for the isotropic cloud and nematic
shadow phases, as visible in Fig. 8. At low polydispersities the
distribution in the nematic shadow phase is very similar to the parent
distribution (albeit with a higher average length). At higher
polydispersities, however, we enter a regime characterized by extreme
fractionation, i.e. the nematic shadow phase is dominated by the longest
rods in the distribution. Remarkably, we do not see a similar transition in
the isotropic shadow phase in Fig. 8, as we did in the Schulz case. This
implies that the fractionation from the nematic cloud phase is much weaker
for log-normal distributions than for Schulz ones.

\begin{figure}
\includegraphics{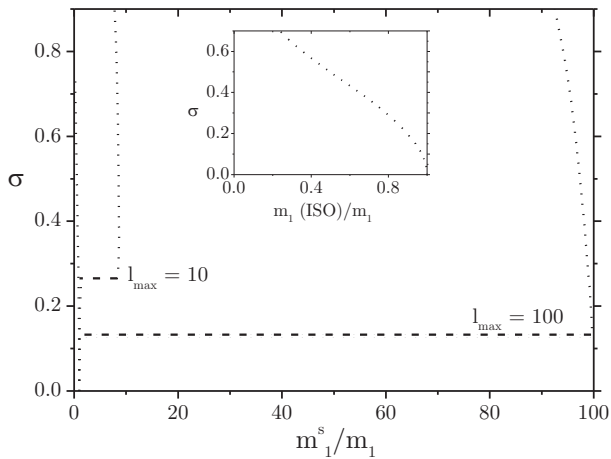}
\caption{Average length $m_{1}^{s}$ in the shadow phases relative
to the parental one $m_{1}$ for log-normal parent distributions with
cutoff lengths $l_{\text{max}}=10$ and $100$ as a function of the parent
polydispersity $\sigma $. The inset shows the relative average length in
the isotropic shadow phase (corresponding to the nematic cloud point). Also here, the results for
$l_{\text{max}}=10$ and $100$ overlap.}
\end{figure}

A crucial difference with the previous results is that the transition
between the regimes occurs discontinuously, that is, at the transition
polydispersities $\sigma _{t}$ the isotropic cloud curves show a kink and
the associated nematic shadow curves exhibit a jump. Precisely at the kink,
the isotropic cloud phase coexists with two different nematic phases, one
containing mostly rods with slightly higher than average length (denoted by $%
N_{I}$) while the second one ($N_{II}$) is dominated by the longest rods in
the distribution. Therefore, this special point marks the beginning of a
stability region for $I-N_{I}-N_{II}$ triphasic equilibria for log-normal
distributions. In Fig. 8 we see that the position of the kink (in terms of $%
\sigma _{t}$) rapidly shifts to lower polydispersities as $l_{\text{max}}$\
increases. From this, we anticipate that the triphasic equilibrium sets in
at almost zero polydispersity (i.e. near monodisperse systems) for very
large cutoff lengths.

\begin{figure}
\includegraphics{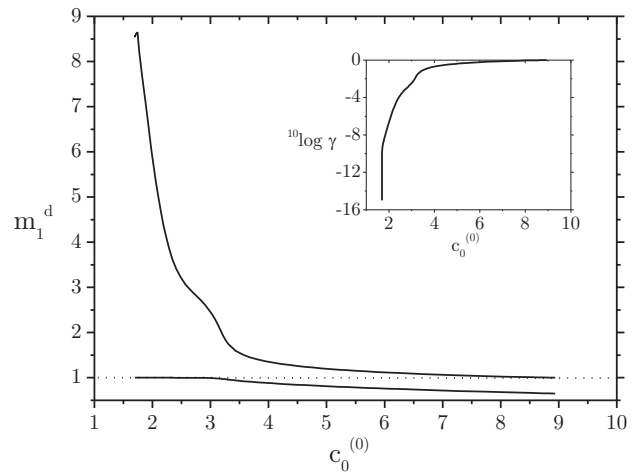}
\caption{Average rod length $m_{1}^{\text{d}}$ in the coexisting
daughter phases (solid lines) plotted versus the concentration $c_{0}^{(0)}$
of the parent across the coexistence region for a log-normal parent  with
$l_{\text{min}}=0.01$ and $l_{\text{max}}=10$ at fixed polydispersity $\sigma=0.4$.
The curve for which $m_{1}^{\text{d}}>1 $ is the nematic branch. The dotted line corresponds to $m_{1}=1$ for
the parent phase. The inset shows the (logarithm of the) fraction $\gamma$ of
the system volume occupied by the nematic phase. Note that the amount of
nematic phase is extremely small in the region $c_{0}^{(0)}<3$.}
\end{figure}
\begin{figure}
\includegraphics{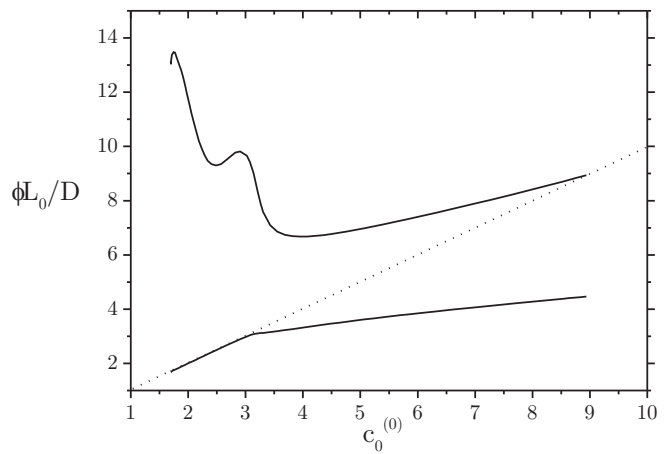}
\caption{Evolution of the scaled volume fraction $\phi L_{0}/D$ in the
coexisting daughter phases across the coexistence region for the same
parent. The nematic branch is the one with the highest volume fraction. The
dilution line is indicated by the dotted line.}
\end{figure}

Like for the Schulz case, we can obtain information about the global phase
behavior for parent distributions at infinite cutoff lengths $l_{\text{max}}$
from the high-cutoff scaling results, shown in Appendix B. The most
important outcome is that the concentrations of the isotropic cloud and
nematic shadow phases go to zero for large cutoff length rather than
approaching asymptotic forms such as in the Schulz case. Furthermore, it is
shown explicitly that the fractionation between the isotropic cloud and
nematic shadow phases is stronger than for the Schulz case.

So far, we have only looked at the onset of phase equilibrium by analyzing
the properties of the cloud and shadow phases. The next step is to explore
the coexistence region in more detail. An intriguing issue is to verify the
region of stability for the isotropic-nematic-nematic triphasic region for
the log-normal case. This will be dealt with in the next section, where we
discuss the phase diagram for a log-normal parent with cutoff length $l_{%
\text{max}}=10$ in more detail.

\section{Inside the I-N coexistence region}

\begin{figure*}
\includegraphics{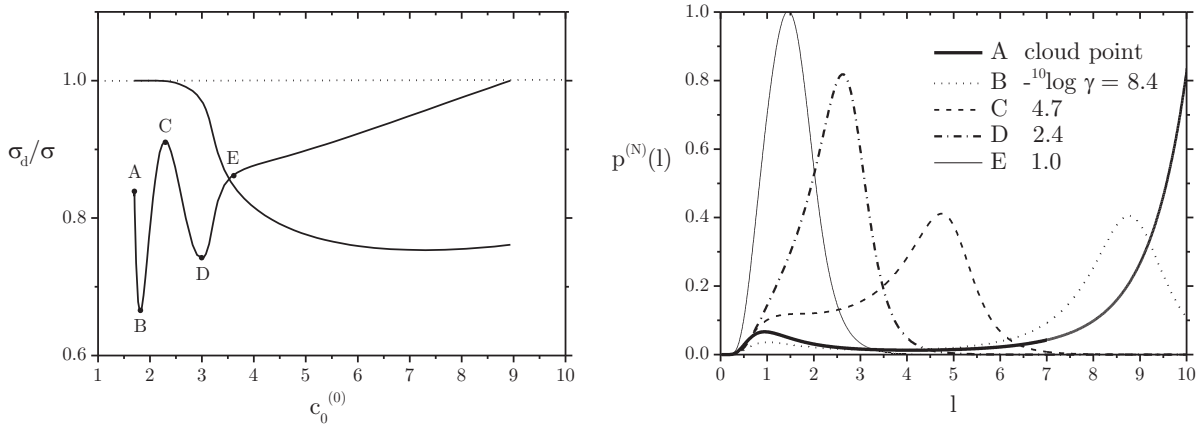}
\caption{(a, left) Relative polydispersity $\sigma _{1}^{\text{d}%
}/\sigma $  of the coexisting daughter phases across the coexistence region
for the same parent. By definition, the dilution line (dotted) is given by $%
\sigma _{1}^{\text{d}}/\sigma =1$.  (b, right) Plot of the normalized length
distributions of the nematic phase across the coexistence region
corresponding to the positions A through E in (a).}
\end{figure*}

Across the coexistence region the equilibium length distributions of the
coexisting phases, which originate from a parent phase with a prescribed
distribution $p^{(0)}(l)$), change continuously as the overall density of
the parent is $c_{0}^{(0)}$ is changed. In the actual calculations however
it is more convenient to impose the fraction $\gamma $ occupied by the
nematic phase rather than $c_{0}^{(0)}$ and calculate the corresponding
densities self-consistently.
In Fig. 9 to 12 we show the properties of the
coexisting isotropic and nematic phases for a log-normal parent distribution
with polydispersity $\sigma =0.4.$  Fig. 9 shows that the average length in
the nematic daughter phase decreases rapidly in the regime $%
c_{0}^{(0)}\lesssim 3$ whereas only weak changes are notable at higher $%
c_{0}^{(0)}$. Furthermore, we see that the volume occupied by the nematic
phase is extremely small in this regime. The same feature is observed in the
volume fraction representation in Fig. 10.In particular, the non-monotonicity of the nematic branch is reflected somewhat clearer here.
The fact that the isotropic branch runs extremely close to the dilution line
for $c_{0}^{(0)}<3$ indicates that the fraction of nematic phase formed must
indeed be very small. The rather exotic oscillations in the behavior of the
polydispersities of the daughter phases in Fig. 11(a) reflect the dramatic
change of the shape of the length distribution in the nematic phase in the
first part ($\gamma <10^{-2}$) of the dilution trajectory as shown in Fig.
11(b). Note that the distribution of the nematic shadow phase is in fact
bimodal, with a small peak around $l=1$ and a much larger one at $l=l_{\text{%
max}}$. When the overall density is increased the second peak shifts to
lower values of $l$ and eventually coincides with the first peak.  When the
overall density has reached about $c_{0}^{(0}=4$ (corresponding to a nematic
phase volume fraction of about 10 \%) the distribution of the nematic phase
resembles the parental one, albeit with a slightly higher average length.

In Fig. 12 we have plotted the evolution of the average length for a parent
with polydispersity $\sigma =0.3$. A peculiar behavior is observed, which is
also reflected in Fig. 13 where the coexistence pressure is plotted versus $%
\gamma $. Clearly, there is a region where the pressure decreases as
function of $\gamma $ which suggests an instability (or van der Waals) loop
indicating a possible destabilization of the nematic phase. In Appendix C we
explicitly show that the local extrema in the osmotic pressure in Fig. 13
represent spinodal points for the nematic phase which indicate that the
coexisting nematic phase indeed becomes locally (and hence globally)
unstable. In the region between the spinodal points (where $\delta (b\beta
\Pi )/\delta \gamma $ is negative), the coexistence between the isotropic
and a single nematic phase also becomes unstable such that a triphasic
isotropic-nematic-nematic ($I-N_{I}-N_{II}$) demixing occurs.

It should be stressed that the actual onset of the three-phase separation is
marked  by {\em binodal} points which we have not located in this study,
except for the kink at $\sigma _{t}$. In general, binodal points are located
at a lower concentration than the spinodal points so that the demixing
usually occurs well before the point where the system becomes locally
unstable. This becomes clear in Fig. 14 where we show the details
of the phase diagram in the vicinity of the kink including the spinodal curves (in terms of ($c_{0}^{(0)}$))
for the coexisting nematic phase.
At the kink $\sigma _{t}=0.264$ the three-phase separation sets in right at
the isotropic cloud point but the spinodal points are located at higher
concentrations. An important feature in Fig. 14 is the presence of a high-$%
\sigma $ consolute (or critical) point at $\sigma =0.373\pm 0.001$ where the
spinodal curves meet. This means that the region of stable triphasic
equilibria does not extend up to large parent polydispersities but closes
off at the consolute point which also constitutes an
endpoint of the binodal curves corresponding to the triphasic equilibria.

Consequently, we can distinguish three regimes in the phase diagram depicted in Fig.  14.
First, below the kink there is a common isotropic-nematic phase separation involving a moderately fractionated
$N_{I}$-phase. Second, in the region between the kink and the consolute
point $0.264\leq \sigma \leq 0.373$ a strongly fractionated $N_{II}$-phase (containing the
longest rods) splits off initially
at the isotropic cloud point. At higher concentrations
(when $I$ and $N_{II}$ coexist in finite amounts) a second nematic $N_{I}$-phase is formed and an $ I-N_{I}-N_{II} $ triphasic equilibium develops.
Upon slightly further concentrating the sample the $N_{II}$-phase disappears and a regular
$I-N_{I}$ biphasic equilibrium is recovered.
The third region is located at parent polydispersities above the
consolute point $\sigma > 0.373 $.
Upon concentrating the isotropic phase a  strongly fractionated nematic phase
(reminiscent of the $N_{II}$ phase) is formed initially  but the composition
of this phase evolves gradually towards a $N_{I}$-type nematic phase as the biphasic region is crossed.

\begin{figure}
\includegraphics{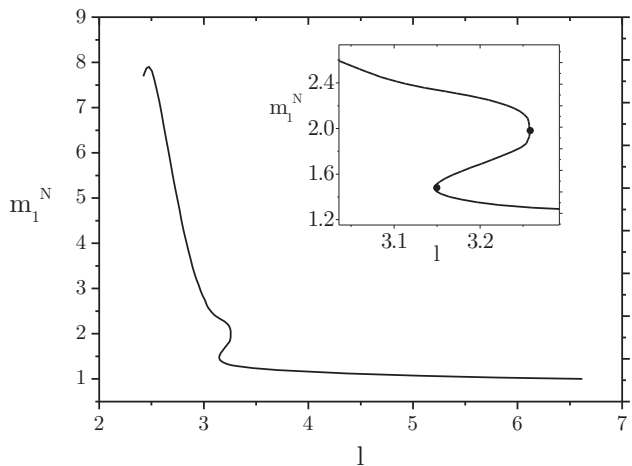}
\caption{Average rod length $m_{1}^{\text{d}}$ in the
nematic phase across the coexistence region for a log-normal parent with the same cutoff lengths
at $
\sigma =0.3$. The inset shows a hysteresis loop indicating that the nematic
phase becomes locally unstable.}
\end{figure}

\begin{figure}
\includegraphics{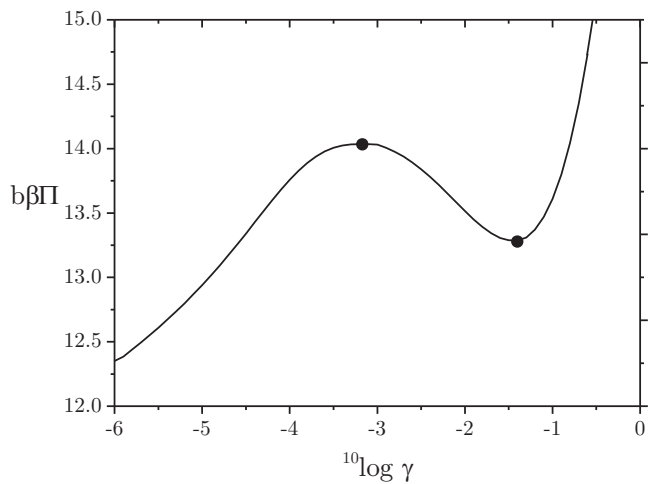}
\caption{Coexistence pressure across the coexistence region
for the same parent. In the region where $\delta (b\beta \Pi )/\delta \gamma
<0$ the nematic phase is locally unstable with respect to a nematic-nematic demixing.}
\end{figure}

\section{Discussion and Conclusions}

\begin{figure}
\includegraphics{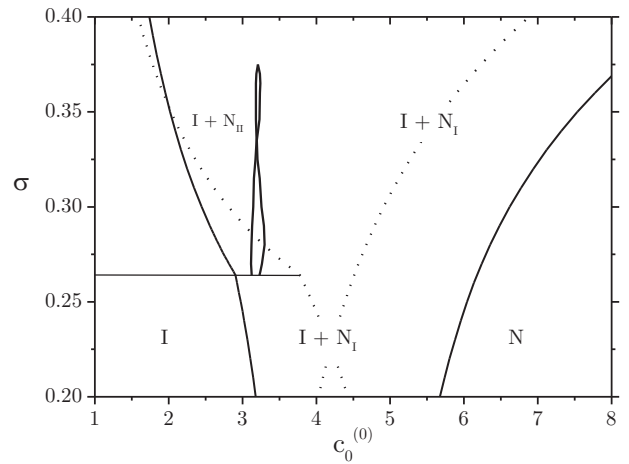}
\caption{Phase diagram for log-normal parent distribution with
cutoff lengths $l_{\min }=0.01$ and $l_{\text{max}}=10$. The thick
curves delimit the spinodal instability region for the coexisting nematic phase
(in terms of the {\em parental} concentration $c_{0}^{(0)}$) indicating a
triphasic $I-N_{I}-N_{II}$ demixing.}
\end{figure}

We have numerically investigated Onsager's second virial theory for
polydisperse hard rods within the Gaussian Ansatz. The onset of
isotropic-nematic phase separation is obtained from the cloud and shadow
curves, which delimit the coexistence region. Within the same numerical
framework, we could also explore the properties of the coexisting phases
across the coexistence region. In this paper, we focussed on systems of
polydisperse hard rods whose lengths can be described by a Schulz or a
log-normal distribution. The basic difference between these two forms is
that the fat-tailed log-normal one contains a significantly higher fraction
of longer rods. For numerical and consistency purposes we truncated the
distributions at both ends at sufficiently low and high cutoff lengths.
Using truncated distributions is also justifiable from an experimental
standpoint. For parent distributions of the Schulz type the phase diagram
contains two fractionation regimes. First, at low parent polydispersities
moderate fractionation occurs and the average rod lengths in the isotropic
and nematic shadow phases are not much different from the average length in
the parental cloud phase. Second, at higher parental $\sigma $ the
fractionation effect is extremely strong and the isotropic and nematic
shadow phases are completely dominated by respectively the longest and the
shortest rods present in the system.

For the exact Onsager model, Speranza and Sollich \cite{sollichonsagerexact}
very recently predicted a kink in the isotropic cloud curve (and a jump in
the corresponding nematic shadow curve) for Schulz parents with cutoff
lengths $l_{\text{max}}$ $>50$. The presence of this kink indicates a region
of stable isotropic-nematic-nematic triphasic equilibria. Here we do not
find any indication for such a three-phase separation for Schulz parents at
least up to $l_{\text{max}}=100$.  The discrepancy may be due to the
Gaussian  Ansatz, which implies that the ODFs are not represented by their
correct equilibrium forms. Moreover, the Schulz form might  be a borderline
case since its tail is too modest to induce a strong demixing but too `fat'
to suppress it completely so that the presence of a kink in the isotropic
cloud curve depends quite sensitively on the precise representation of the
ODF.

Although the Gaussian ODF is not a solution of the exact stationarity
condition for the ODF, it does satisfy the exact high-density scaling
relation \cite{vanroijmulderscaling}. This means that the properties of
highly ordered nematic states are described very well by the Gaussian form.
In fact, the description becomes {\em exact} for infinitely aligned states.
A manifestation of this is the osmotic pressure for the nematic phase, Eq. (%
\ref{driecee}), which is the {\em exact} high-density result \cite
{vanroijmulderscaling}.  Consequently,  for our polydisperse systems, we
expect the Gaussian Ansatz to work increasingly well both for highly
concentrated nematic phases and nematics that are dominated by the longest
rods. In both cases, the nematic alignment of all species is expected to be
very pronounced such that the use of the scaling ODF (for all $l$) is
justified. To verify this notion, we have plotted the variational
parameter as a function of length, for both the nematic shadow phase and the
nematic cloud phase corresponding to a Schulz parent in Fig. 15. Since the
Gaussian ODF is expected to be the least correct for the shortest rods
(which show the weakest alignment), we focus on the interval $l_{\text{max}%
}<1$. In order for the results to be self-consistent, the alignment must be
strong enough and hence the variational parameter must be sufficiently large
(say $^{10}\log \alpha >1$) for all lengths. Fig. 15 (a) shows that this is
not entirely the case; in the regime of low fractionation the shortest rods
(with $l\lesssim 0.4$) are not sufficiently aligned by the longer rods so
that the Gaussian description fails here. In the regime $\sigma >0.5$, which
is physically the most interesting one, the ordering of the short rods is
much higher due to the presence of very long rods in the nematic shadow
phase, and the Gaussian Ansatz is fully justified. Similarly for the nematic
cloud phase in Fig. 15 (b) we see that the shortest rods are not well
represented by the Gaussian ODF at low $\sigma $ but much better at $\sigma
>\sigma _{t}$ where the variational parameter increases several orders of
magnitude due to a dramatic increase of the concentration of the nematic
cloud phase (see also Fig. 2). Obviously, for any rod length significantly
larger than $l_{\text{min }}$ the Gaussian ODF works very well because $%
\alpha $ generally becomes extremely large for any $\sigma $. Therefore we
conclude that given the fact that the composition of the nematic phases is
dominated by the longest rods, particularly in the physically relevant
cutoff-dependent regime, the Gaussian description is an appropriate tool in
our study.

\begin{figure}
\includegraphics{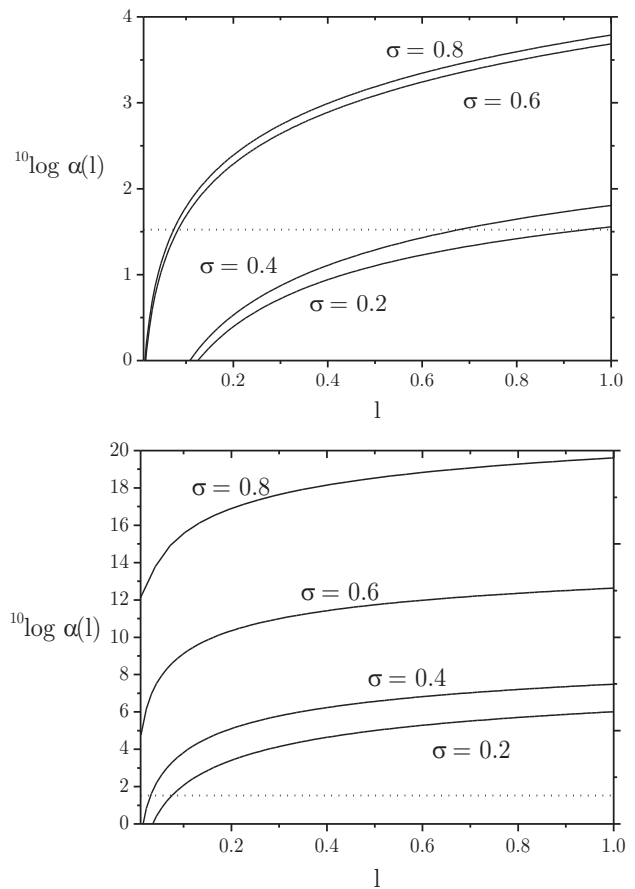}
\caption{Behavior of the Gaussian variational
parameter $\alpha(l)$ for the shortest rods of a Schulz
parent with $l_{\text{min}}=0.01$ and $l_{\text{max}}=100$ (see also Fig. 1).
(a, top) Results for the nematic shadow phase at various $\sigma$. (b, bottom)
Same for the nematic cloud phase. For comparison the result for the monodisperse
system \protect\cite{Vroege92} ($\alpha=33.4$) is indicated by the dotted line.}
\end{figure}

We now turn to the log-normal case. The fractionation scenario we observe
there is qualitatively the same as the one for the Schulz case; weak
fractionation occurs at low $\sigma $ where the distributions in the shadow
phases are reminiscent of the parental one, but dramatic segregation effects
take place above some threshold $\sigma ,$ particularly between the
isotropic cloud and the nematic shadow phase. A crucial difference with the
results for the Schulz case however is that the transition between the two
fractionation regions shows a discontinuity at some $\sigma =\sigma _{t}$.
At this point, the isotropic cloud curve shows a kink which corresponds to a
jump in the nematic shadow phase. The jump indicates that an isotropic cloud
phase must coexist with two different shadow phases, one containing mostly
rods of slightly higher than average length (the $N_{I}$-phase) and the
other one predominantly containing very long rods (the $N_{II}$-phase). The
kink also marks the beginning of a region of stable $I-N_{I}-N_{II}$
triphasic equilibria. For log-normal distributions with a moderate
cutoff-length $l_{\text{max}}=10$ we found that the triphasic region opens
up at a fairly low polydispersity $\sigma _{t}=0.264.$ This value will
decrease for larger cutoff lengths and eventually go to zero when $l_{\text{%
max}}$ approaches infinity. This indicates that adding a very small
fraction of long rods to a weakly polydisperse system of much shorter rods
can already induce a three-phase demixing. A similar effect is observed in
binary mixtures of long and short hard rods with sufficiently large length
ratios \cite{LekkerVroeg}.

By numerically analyzing the spinodal instability criteria for the nematic
phase across the coexistence region for the case $l_{\text{max}}=10$  we
have shown that the triphasic area does not extend up to very large parent
polydispersities but terminates at a consolute point located at $\sigma
=0.373\pm 0.001.$ This means that isotropic-nematic-nematic triphasic
equilibria are only expected to occur in a small interval of parent
polydispersities, namely $0.264\leq \sigma \leq 0.373$. However, it should
be mentioned that we have not been able to determine the corresponding
binodal points, which mark the actual onset of three-phase separation from
the isotropic-nematic biphasic equilibria. Considering the results for the
approximate Onsager ${\cal P}_{2}$-model, as numerically analyzed by
Speranza and Sollich \cite{sollichonsagerP2fattail}, we expect these
triphasic equilibria to be limited to a very small density interval across
the coexistence region, which makes it very hard to observe the phenomenon
in experiment. Another problem is that the fraction of the system volume
occupied by the nematic phases is predicted to be at the most 0.1 \%, so
that it will be very difficult to distinguish (or even detect) the two
different nematics. Therefore we must conclude that, although the log-normal
distribution contains sufficiently long rods to induce a demixing of the
nematic phase, the (mole) fraction of these rods is too small to give rise
to an observable fraction of the demixed nematic phases.

Experimentally, phenomena such as a  broadening of the biphasic region
and a fractionation effect have been observed unequivocally in a number of experimental
studies \cite{fradenmaret,donkai93,Buining93,Itou}.
Observations of triphasic isotropic-nematic-nematic
equilibria  were however only reported for systems whose length
distributions appear to be more or less bimodal rather than unimodal.
These bimodal shapes were either accomplished deliberately by mixing species
with different lengths, as done by Itou \cite{Itou} with semiflexible schyzophyllan rods
or caused by the presence of large aggregates as found  by Kajiwara {\em et al.} \cite{Kajiwara} in systems of rigid imogolite rods.
Buining {\em et al.} \cite{Buining93}
observed the formation of second nematic phase in systems of polymer-coated (hard) boehmite rods,
albeit a long time after the two-phase isotropic-nematic phase separation had finished.
Also there, the triphasic demixing is probably due to the presence of a small number of very long rods or
aggregates present in the system causing the length distribution to be (slightly) bimodal \cite{Buiningbeharing}.

The experimental results therefore seem to indicate that having a three-phase separation
with observable fractions of all phases requires some degree of bimodality (with a sufficiently large length ratio between the short and long species)
in the parental length distribution.  Hence, an intriguing
issue left open for future investigation is the question
how the triphasic phenomenon predicted for the log-normal distributions
change for a parent distribution with a slightly bimodal shape (e.g.
with a small second peak just below $l_{\max }$). In particular, a bimodal distribution may give rise
to enhanced fractionation behavior and more pronounced triphasic equilibria than predicted for the unimodal log-normal
form.

\section*{Appendix A: Numerical procedure}

The self-consistency equations for $\tilde{\alpha}(l)$ and $\mu _{\text{ex}%
}^{(N)}(l),$ Eqs. (\ref{selfconsistency}) and (\ref{haa}), were solved using
a numerical grid of lengths. The iterative scheme we used is analogous to
the one described by Herzfeld {\em et al.} \cite{herzfeldgrid} for computing
the numerically exact equilibrium ODF\ of the monodisperse Onsager model.
The $l-$interval $[l_{\text{min}},l_{\text{max}}]$ was discretized into $N$
(not necessarily equal) parts. The mesh size must be chosen very carefully,
particularly for large $l_{\max }$, because the distributions become
considerably peaked at low polydispersities. Therefore, for parent
polydispersities lower than approximately $\sigma =0.25$ we chose to divide
the integration interval into three regimes. The interval $[1-8\sigma
^{2},1+8\sigma ^{2}]$ in the vicinity of the peak, where the distribution
changes rapidly, was discretized into $\frac{3}{5}N$ equal parts, and the
intervals $[l_{\min },1-8\sigma ^{2}]$ and $[1+8\sigma ^{2},l_{\text{max}}],$
where the distribution is generally much smoother, were both discretized
into $\frac{1}{5}N$ equal parts. For parent polydispersities larger than $%
\sigma =0.25$ the entire interval was discretized into $N$ equal parts. It
has been proved sufficient to use $N=150$ in order to obtain quantitatively
reliable results. However, for the calulation of the full phase split (Sec.
III-C) a smaller mesh size ($N=50)$ was used to limit the computational
burden. Increasing the number of mesh points leads to only marginally
different results in this case while the calculation time increased
dramatically. For small polydispersities we took Eqs. (\ref{alphanarrow})
and (\ref{muexnarrow}) as initial guesses and the successive iteration was
performed until the following convergence criteria were satisfied
\begin{eqnarray}
\max_{n=1,...N}\left| \tilde{\alpha}(l_{n+1})-\tilde{\alpha}(l_{n})\right|
&<&10^{-6},  \nonumber \\
\max_{n=1,...N}\left| \tilde{\mu}_{\text{ex}}^{(N)}(l_{n+1})-\tilde{\mu}_{%
\text{ex}}^{(N)}(l_{n})\right| &<&10^{-6}.
\end{eqnarray}
After each iteration step 10\% of the new solution had to be added to 90 \%
of the previous one for the next iteration step to ensure the convergence of
the method.

The iteration algorithm we figured out to calculate the phase equilibria, in
particular for the isotropic cloud and shadow phases and the full phase
split, can be described as follows. First, the corresponding equilibrium
forms for $\tilde{\alpha}(l)$ and $\tilde{\mu}_{\text{ex}}^{(N)}(l)$ were
calculated for a given set of starting concentrations. These results were
then put into the self-consistency equations for the cloud and shadow
concentrations to obtain new values. These equations were obtained by
recasting the self-consistency conditions for the moments (e.g. Eq. (\ref
{normcond})) in an iterative form such that the concentrations could be
calculated by a simple fixed-point iteration.

Finally, for the new concentrations, corresponding forms for $\tilde{\alpha}%
(l)$ and $\tilde{\mu}_{\text{ex}}^{(N)}(l)$ were computed and substituted
again into the self-consistency equations. This procedure was repeated until
all concentrations had converged to within 10$^{-5}$. To ensure convergence
of this iteration loop a damping percentage of 80 \% was used, which means
that only 20 \% of the change was retained at each iteration step.

\section*{Appendix B: High-cutoff scaling results}

In this appendix we focus on the properties of the nematic shadow phase in
the cutoff-dependent regime ($\sigma >\sigma _{t}$) for systems with
infinite cutoff lengths. For the exact Onsager model, Speranza and Sollich
\cite{sollichonsagerexact} made a detailed analysis of these properties
based upon the high-density (and high-cutoff) scaling forms of the exact
nematic ODF. \ Here, we will not reproduce the analysis but merely show that
the scaling form for the length distribution in the nematic shadow phase in
the limit $l_{\text{max}}\rightarrow \infty $ is {\em analogous} to the one
obtained in Ref. \cite{sollichonsagerexact}. Consequently, all scaling
properties which follow from the Gaussian approximation must be {\em exactly
the same} as the ones derived from the exact high-cutoff scaling results.

The first step is to solve the coupled set of consistency equations, Eqs. (%
\ref{selfconsistency}) and (\ref{haa}). In order to obtain analytic
solutions for these nonlinear integral equations we exploit the fact that
the nematic shadow phase is completely dominated by the longest rods in the
system at $\sigma >\sigma _{t}$. \ When the cutoff length increases the
length distribution in the nematic shadow will be more and more peaked at $%
l=l_{\text{max}}$. In the limit of infinite $l_{\text{max}}$ it is therefore
justified to use the Ansatz $p^{(N)}(l)=\delta (l-l_{\text{max}})$ which
suggests an effectively monodisperse nematic shadow phase only containing
the longest rods in the system. Substituting the delta-function in Eqs. (\ref
{selfconsistency}) and (\ref{haa}) allows us to obtain asymptotic forms for
the Gaussian variational parameter $\tilde{\alpha}(l)$ and the excess
chemical potential \ $\tilde{\mu}_{\text{ex}}^{(N)}(l)$ of the nematic
shadow phase. These expressions now read
\begin{eqnarray}
\tilde{\alpha}(l) &=&l_{\text{max}}^{4}{\cal F}(l/l_{\text{max}}),
\label{asympformalfa} \\
\tilde{\mu}_{\text{ex}}^{(N)}(l) &=&\frac{l}{l_{\text{max}}}2^{3/2}\sqrt{1+%
{\cal F}^{-1}(l/l_{\text{max}})},  \label{asympformmu}
\end{eqnarray}
where ${\cal F}$ is given by
\begin{equation}
{\cal F}(l/l_{\text{max}})=\frac{1}{2}\left[ \sqrt{1+8\left( \frac{l}{l_{%
\text{max}}}\right) ^{2}}-1\right].
\end{equation}
Note that ${\cal F}(l/l_{\text{max}})$ always has a value between zero and
unity. A close inspection of Eqs. (\ref{asympformalfa}) and (\ref
{asympformmu}) reveals that the variational parameter $\tilde{\alpha}(l)$
scales as $\tilde{\alpha}(l)\propto l_{\text{max}}^{4}$ whereas the excess
chemical potential $\tilde{\mu}_{\text{ex}}^{N}(l)$ remains of the order $%
{\cal O}(1)$ for all lengths. \ Using this in Eq. (\ref{exprnemshadow}) we
can write down a scaling expression for the length distribution in the
nematic shadow, which in its general form reads
\begin{equation}
p^{(N)}(l)=\text{cst}\frac{c_{0}^{(0)}}{(c_{0}^{(N)})^{3}}l_{\text{max}%
}^{-4}p^{(0)}(l)\exp \left[ 2c_{0}^{(0)}l-{\cal W}(l/l_{\text{max}})\right] ,
\label{scaling shadow}
\end{equation}
where ${\cal W}$ is a contribution of the order ${\cal O}(1)$:
\begin{equation}
{\cal W}(l/l_{\text{max}})=\ln {\cal F}(l/l_{\text{max}})+\tilde{\mu%
}_{\text{ex}}^{(N)}(l/l_{\text{max}}).
\end{equation}
Note that ${\cal W}$ attains its maximum ${\cal W}=4$ for $l=l_{\text{max}}$.
The scaling solution Eq. (\ref{scaling shadow}) is completely analogous to
the one found in Ref. \cite{sollichonsagerexact}, the only differences being
the exact form of ${\cal W}(l/l_{\text{max}})$ and the constant cst.
However, since these contributions are both subleading in the limit $l_{%
\text{max}}\rightarrow \infty $ they are irrelevant for the rest of the
analysis and hence do not influence the scaling results. The similarity
between the exact high density scaling analysis and the Gaussian
approximation is also confirmed by the fact that both theories predict the
same high-density scaling result for the nematic osmotic pressure, namely $%
b\beta \Pi =3c_{0}^{(N)}$.

For the sake of completeness, let us now briefly outline the basic results
of the large cutoff scaling analysis. For a comprehensive treatment of this
subject the reader is referred to Ref. \cite{sollichonsagerexact}. For a
Schulz parent, we may use Eq. (\ref{schulz}) in Eq. (\ref{scaling shadow})
to obtain
\begin{equation}
p^{(N)}(l)={\cal K}_{N}l_{\text{max}}^{-4}l^{z}\exp \left[ \varepsilon l-%
{\cal W}(l/l_{\text{max}})\right] ,  \label{scalingschulz}
\end{equation}
where $\varepsilon =2c_{0}^{(0)}-(z+1)$ and $z=\sigma ^{-2}-1.$ For very
large $l$ the exponent $\exp [\varepsilon l]$ will be the dominating
contribution. At $\sigma >\sigma _{t},$ the nematic shadow is supposed to be
dominated by the longest rods and the distribution $p^{(N)}(l)$ should
therefore be an increasing function of length. This requires $\varepsilon $
to be positive and yields the condition $c_{0}^{(0)}>\frac{1}{2}(z+1)$.
Since the concentration of the cloud phase appears to decrease with
increasing $l_{\text{max}}$ this then implies that the isotropic cloud curve
(and hence the nematic shadow curve) has a finite lower bound for large
cutoff lengths. These limiting solutions, for which $\varepsilon =0$,
therefore read
\begin{eqnarray}
c_{0}^{(0)} &=&\frac{1}{2\sigma ^{2}},  \nonumber \\
c_{0}^{(N)} &=&\frac{1}{6\sigma ^{2}}\left( 1+\frac{1}{2\sigma ^{2}}\right) ,
\label{limitingcurves}
\end{eqnarray}
using Eq.(\ref{condeqpressure}), with $c_{1}^{(0)}=c_{0}^{(0)}$. These
results are plotted in Fig. 1. To be consistent, let us now look for a
solution for the transition polydispersity $\sigma _{t}$ above which the
nematic shadow phase for a Schulz distributed parent is completely dominated
by the longest rods. We start with the concentration of the nematic shadow
phase which is proportional to the integral over the normalized length
distribution, i.e $c_{0}^{(N)}\propto \int p^{(N)}(l)dl$. From Eq. (\ref
{scalingschulz}) we thus obtain
\begin{equation}
c_{0}^{(N)}\propto l_{\text{max}}^{-4}\int_{0}^{l_{\text{max}}}l^{z}\exp %
\left[ \varepsilon l-{\cal W}(l/l_{\text{max}})\right] dl.
\end{equation}
For the sake of convenience we have set $l_{\text{min}}$ equal to zero.
Since the integrand is dominated by the exponent $\exp [\varepsilon l]$ for
large $l$ we may approximate the integral by bringing all slowly varying
contributions in front of the integral sign and evaluating them at $l=l_{%
\text{max}}$, which gives
\begin{eqnarray}
c_{0}^{(N)} &\propto &l_{\text{max}}^{-4}\left| l^{z}\exp [-{\cal W}(l/l_{%
\text{max}})]\right| _{l=l_{\text{max}}}\int_{0}^{l_{\text{max}}}dl\exp
[\varepsilon l]  \nonumber \\
c_{0}^{(N)} &\propto &l_{\text{max}}^{z-3}\frac{\exp [\varepsilon l_{\text{%
max}}]}{\varepsilon l_{\text{max}}}.  \label{scalingc}
\end{eqnarray}
The next step is to recast the latter equation into a scaling relation for $%
\varepsilon .$ Taking the logarithm on both sides of Eq. (\ref{scalingc})
gives
\begin{equation}
\varepsilon \propto (3-z)\frac{\ln l_{\text{max}}}{l_{\text{max}}}+\frac{\ln
\varepsilon l_{\text{max}}}{l_{\text{max}}}+{\cal O}(l_{\text{max}}^{-1}).
\label{epsilon}
\end{equation}
From the known limits $\varepsilon l_{\text{max}}\rightarrow \infty $ and $%
\varepsilon \downarrow 0$ for $l_{\text{max}}\rightarrow \infty $ we can
deduce that $\varepsilon l_{\text{max}}$ must increase more slowly than
linearly with $l_{\text{max}}$. Consequently, the second and third terms in
Eq. (\ref{epsilon}) are both subleading contributions so that we retain up
to leading order
\begin{equation}
\varepsilon \propto (3-z)\frac{\ln l_{\text{max}}}{l_{\text{max}}},
\label{epsilonfinal}
\end{equation}
which shows that $\varepsilon l_{\text{max}}$ indeed increases
logarithmically rather than linearly as we already anticipated. However, in
order to make this result fully self-consistent it is also required that $z$
$<3$ (and correspondingly $\sigma >0.5$) since $\varepsilon $ must be
positive. This means that $\sigma =0.5$ is a lower bound for the
cutoff-dependent regime in the limit $l_{\text{max}}\rightarrow \infty $. In
other words, the transition from the low fractionation regime to the regime
where the nematic shadow is completely dominated by the longest rods occurs
exactly at $\sigma _{t}=0.5$ for Schulz parents with infinitely high cutoff
lengths.

We now turn to the average length in the nematic shadow phase which is
related to the first moment density $c_{1}^{(N)}\propto \int lp^{(N)}(l)dl$.
Analogously to Eq. (\ref{scalingc}) it follows that $c_{1}^{(N)}\propto $ $%
l_{\text{max}}^{z-2}\exp [\varepsilon l_{\text{max}}]/\varepsilon l_{\text{%
max}}$ and that the average length hence scales as $\left\langle
l\right\rangle \equiv $ $c_{1}^{(N)}/c_{0}^{(N)}\propto l_{\text{max}}$.
Since the distribution in the nematic shadow phase is dominated by $\exp
[\varepsilon l]$ we expect that only rods whose lengths are of the order $%
{\cal O}(1/\varepsilon )$ smaller than $l_{\text{max}}$ contribute to the
average length. We therefore can write
\begin{equation}
\left\langle l\right\rangle =l_{\text{max}}\left[ 1-{\cal O}\left( \frac{1}{%
\varepsilon l_{\text{max}}}\right) \right] ,
\end{equation}
and from Eq. (\ref{epsilonfinal})
\begin{equation}
\left\langle l\right\rangle =l_{\text{max}}\left[ 1-{\cal O}\left( \frac{1}{%
\ln l_{\text{max}}}\right) \right] .  \label{laverageschulz}
\end{equation}
This result shows that the average length in the nematic shadow in principle
diverges for $l_{\text{max}}\rightarrow \infty $ but the logarithmic
correction causes the actual $\left\langle l\right\rangle $ to be
significantly lower than $l_{\text{max}}$.

A similar treatment can be given for a log-normal parent
distribution.  However, the analysis for the log-normal case is even more
involved and we will only present the basic results and refer to Ref. \cite
{sollichonsagerexact} for details. First, the concentration of the
isotropic cloud phase appears to have the following $l_{\text{max}}$%
-dependence
\begin{equation}
c_{0}^{(0)}=\frac{\ln ^{2}l_{\text{max}}}{4\ln (1+\sigma ^{2})l_{\text{max}}}%
+{\cal O}\left( \frac{\ln l_{\text{max}}}{l_{\text{max}}}\right) ,
\label{cloudscalelog}
\end{equation}
which is crucially different from the Schulz case because the concentration
of the cloud and shadow phases now tend to zero rather than approaching
boundary values as in the Schulz case. Second, the average length can be
shown to behave as
\begin{equation}
\left\langle l\right\rangle =l_{\text{max}}\left[ 1-{\cal O}\left( \frac{1}{%
\ln ^{2}l_{\text{max}}}\right) \right] .
\end{equation}
Similarly to the Schulz distribution, the average length scales as $%
\left\langle l\right\rangle \propto l_{\text{max}}$ but the correction term
is now considerably smaller which implies that the fractionation effect is
much more pronounced for log-normal distributions at $\sigma >\sigma _{t}$,
as we already noticed by comparing Figs. 3 and 8.

\section*{Appendix C: Polydisperse spinodal instability criterion for the
nematic phase}

In this appendix we show that the anomalous behavior of the coexistence
pressure in Fig. (13) can be related to the local instability of the
coexisting nematic phase. In particular, it is established that the (local)
extrema in the coexistence pressure are directly related to the spinodal
points where the nematic phase becomes locally (and hence globally)
unstable. Let us first consider the Gibbs-Duhem equation, which in its
general form is given by
\begin{equation}
\left( \frac{\partial P}{\partial \rho _{j}}\right) _{T, \rho_{i\neq j}}=\sum_{i}\rho
_{i}\left( \frac{\partial \mu _{i}}{\partial \rho _{j}}\right) _{T, \rho_{i\neq j}},
\end{equation}
with $\rho _{i}=N_{i}/V$. For a continuous density distribution $c(l)$ this
relation can be expressed in terms of functional derivatives
\begin{equation}
\frac{\delta (b\beta \Pi )}{\delta c(l^{\prime })}=\int dlc(l)\frac{\delta
\beta \mu (l)}{\delta c(l^{\prime })}.  \label{gibbsduhem}
\end{equation}
We now focus on the evolution of the length distribution of the nematic
phase (in coexistence with an isotropic phase) across the coexistence region
and denote this distribution by $c^{\ast }(l)$. The shape of $c^{\ast }(l)$
depends uniquely on the nematic fraction $\gamma $, so that the curves in
Figs. 12 and 13 represent trajectories parametrized by $\gamma $. As to the
osmotic pressure, let us now consider an infinitesimal change of the
coexistence pressure $\delta (b\beta \Pi )^{\ast }$ corresponding to an
infinitesimally small dispacement $\delta c^{\ast }(l)$ on the trajectory in
Fig. (13). Using the Gibbs-Duhem relation Eq. (\ref{gibbsduhem}), it then
follows that
\begin{eqnarray}
\delta (b\beta \Pi )^{\ast } &=&\int dl^{\prime }\left. \frac{\delta (b\beta
\Pi )}{\delta c(l^{\prime })}\right| _{c^{\ast }(l^{\prime })}\delta c^{\ast
}(l^{\prime })  \nonumber \\
&=&\int dlc^{\ast }(l)\int dl^{\prime }\left. \frac{\delta \beta \mu (l)}{%
\delta c(l^{\prime })}\right| _{c^{\ast }(l^{\prime })}\delta c^{\ast
}(l^{\prime }).
\end{eqnarray}
Since $c^{\ast }(l)>0$ for all $l$ the infinitesimal pressure change is only
zero when
\begin{equation}
\int dl^{\prime }\left. \frac{\delta ^{2}f}{\delta c(l)\delta c(l^{\prime })}%
\right| _{c^{\ast }(l)}\delta c^{\ast }(l)=0,
\end{equation}
which is precisely the spinodal instability criterion for polydisperse
systems \cite{sollichwarrenmomenttheory}, albeit restricted to composition
fluctuations $\delta c^{\ast }(l)$ which lie {\em on the trajectories}
parametrized by $\gamma $. This result therefore shows that the local
extrema in the coexisting pressure where $\delta (b\beta \Pi )^{\ast }=0$
represent spinodal points indicating local instability of the coexisting
nematic phase. The instability direction $\delta c^{\ast }(l)$ is defined as
the change of the nematic length distribution $c(l)$ corresponding to an
infinitesimal displacement $\delta \gamma $ on the trajectory in Fig. 13.
The presence of the spinodal points for the nematic phase imply the
existence of a isotropic-nematic-nematic triphasic region.

\end{document}